\documentclass{aa}

\usepackage{graphicx}
\usepackage{amsmath}
\usepackage{amssymb}
\usepackage{import}
\usepackage{xcolor}
\usepackage{ulem}
\usepackage[bookmarks=true,
  colorlinks,
  linkcolor=blue,
  urlcolor=blue,
  citecolor=blue,
  plainpages=false,
  pdfpagelabels,
  final,
  breaklinks=true]{hyperref}
\usepackage[capitalise]{cleveref}
\crefname{section}{Sect.}{Sects.}
\usepackage{scalerel,stackengine}
\usepackage{xfrac}

\bibpunct{(}{)}{;}{a}{}{,} 
\usepackage[varg]{txfonts}

\newcommand{\Mpch}{\ensuremath{h^{-1}\,\mathrm{Mpc}}}
\newcommand{\hMpc}{\ensuremath{h\,\mathrm{Mpc}^{-1}}}
\newcommand{\hMsolar}{\ensuremath{h^{-1}\,M_\odot}}

\newcommand{\calP}{\mathcal{P}}
\newcommand{\PT}{\texttt{PineTree} }
\newcommand{\Nmix}{\ensuremath{N_{\rm{mix}}}}

\begin{document}

\title{\texttt{PineTree}: A generative, fast, and differentiable halo model for wide-field galaxy surveys}
\subtitle{}

\author{Simon Ding\inst{1}\fnmsep\thanks{ Corresponding author.}
   \and Guilhem Lavaux\inst{1}
   \and Jens Jasche\inst{2}}

\institute{Sorbonne Université, CNRS, UMR 7095, Institut d’Astrophysique de Paris, 98 bis boulevard Arago, 75014 Paris, France \\
           \email{simon.ding@iap.fr}
      \and The Oskar Klein Centre, Department of Physics, Stockholm University, Albanova University Center, SE 106 91 Stockholm, Sweden}

\date{Received XXX / Accepted YYY}

\abstract 
{
Accurate mock halo catalogues are indispensable data products for developing and validating cosmological inference pipelines.
A major challenge in generating mock catalogues is modelling the halo or galaxy bias, which is the mapping from matter density to dark matter halos or observable galaxies.
To this end, $N$-body codes produce state-of-the-art catalogues.
However, generating large numbers of these $N$-body simulations for big volumes, especially if magnetohydrodynamics are included, requires significant computational time.
}
{
We introduce and benchmark a differentiable and physics-informed neural network that can generate mock halo catalogues of comparable quality to those obtained from full $N$-body codes.
The model design is computationally efficient for the training procedure and the production of large mock catalogue suites.
} 
{
We present a neural network, relying only on \(18\) to \(34\) trainable parameters, that produces halo catalogues from dark matter overdensity fields.
The reduction of network weights is realised through incorporating symmetries motivated by first principles into our model architecture.
We train our model using dark matter only $N$-body simulations across different resolutions, redshifts, and mass bins.
We validate the final mock catalogues by comparing them to \(N\)-body halo catalogues using different \(N\)-point correlation functions.
}
{
Our model produces mock halo catalogues consistent with the reference simulations, showing that this novel network is a promising way to generate mock data for upcoming wide-field surveys due to its computational efficiency.
Moreover, we find that the network can be trained on approximate overdensity fields to reduce the computational cost further.
We also present how the trained network parameters can be interpreted to give insights into the physics of structure formation.
Finally, we discuss the current limitations of our model as well as more general requirements and pitfalls for approximate halo mock generation that became evident from this study.
}
{}

\keywords{
	methods: statistical -- dark matter -- galaxies: halos -- galaxies: abundances -- galaxies: statistics -- large-scale structure of Universe
}

\titlerunning{}
\authorrunning{} 
\maketitle



\section{Introduction}
\label{sec:intro}

Modern surveys like \textit{Euclid} \citep{euclid, euclid2}, DESI \citep{desi}, LSST \citep{lsst}, PFS \citep{pfs}, SKA \citep{ska1, ska2}, WFIRST \citep{wfirst} and SPHEREx \citep{spherex} are transforming cosmology into a data-driven research field with their unprecedented survey volume and instrument sensitivity.
The constraints on cosmological parameters will shift from a noise-dominated regime to a systematics-dominated one.
Thus, progress in the field will critically depend on our ability to model the data with ever-increasing accuracy and precision.

To this end, simulated observations or mock data are essential data products for developing new models and cosmological inference pipelines, as demonstrated by \citet{mock_challenge}.
Another application utilises mocks to estimate the sample covariance matrix numerically for clustering statistics with the two-point and three-point correlation functions (like \citealt{numerical_covariance_for_boss}).
With the advance of machine learning techniques in recent years, cosmologists have also explored various applications by leveraging simulations to train deep neural networks \citep[e.g.][]{cosmology_with_ML, cosmology_with_ML4, cosmology_with_ML2, cosmology_with_ML3, camels, lyman_alpha_nn, map2map_emulator, legin2024posterior}.
Notably, the combination of machine learning with Bayesian statistics known as simulation-based inference (SBI) or implicit likelihood inference (ILI) has established itself as a promising framework in cosmological analysis \citep{simbig, ltu-ili, makinen2022cosmic, ili_with_galaxies}.

The state-of-art method for generating mock observables involves \(N\)-body codes such as \texttt{Gadget} \citep{gadget4}, \texttt{Ramses} \citep{ramses}, \texttt{ENZO}, \texttt{Pkdgrav-3} (an earlier version described by \citealt{pkdgrav}) and \texttt{Abacus} to trace the evolution of matter over cosmic time.
Subsequently, halo finders like Friends-of-friends \citep{fof}, \texttt{SUBFIND} \citep{subfind}, \texttt{HOP} \citep{hop}, \texttt{AdaptaHOP} \citep{adaptaHOP,adaptahop_paper}, \texttt{Amiga Halo Finder} \citep{ahf}, \texttt{ASOHF} \citep{asohf}
 and {\texttt ROCKSTAR} \citep{rockstar} group the particles hierarchically into halos or galaxies.
These powerful \(N\)-body simulations can be computationally expensive, especially if they aim to cover large survey volumes, include baryonic physics, or explore the effect of different astrophysical and cosmological parameters like \citet{EuclidSkyFlagship, quijote, tng, abacus, uchuu}.
Running these simulations requires millions of CPU hours and petabytes of storage.
A growing need for simulation suites with large sets of independent samples aggravates the issue of computational resources.
Work done by \citet{covariance_samples_size} shows that more than \(5000\) independent mock observations are necessary in a Euclid-like setting to estimate the numerical covariance with adequate accuracy for the matter power spectrum.
Similarly, the performance of machine learning methods is directly dependent on the number of available simulations.
In situations with sparse training data, deep neural networks are likely to overfit and thus may yield biased or overconfident results, as discussed by \citet{ili-training} for ILI algorithms.

Several methods have been proposed to alleviate the reliance on running large sets of high fidelity \(N\)-body simulations.
For example, techniques like CARPool \citep{carpool1, carpool2} or ensembling \citep{ili-training} reduce the required number of expensive \(N\)-body runs.
An alternative approach involves building approximate simulators that are computationally faster than \(N\)-body codes while maintaining sufficient accuracy.
\Citet{covariance_samples_size} demonstrates that for matter power spectrum analysis in a Euclid-like scenario, the cosmological constraints are substantially weaker with scale cuts \(k \leq 0.2\,\hMpc\) whereas only adopting linear covariance to include smaller scales results in overconfident parameter estimations.
This effect will be exacerbated for other summary statistics such as \(k\)-th nearest neighbour \citep{knn}, density-split clustering \citep{dsc}, skew spectrum \citep{skew_spectra}, wavelet scattering transforms \citep{wavelet1, wavelet2}, cosmic voids \citep{void1, void2}, or full field-level methods \citep{borg1, field_level_wl, field_level_bao_marginalization, field_level_cmb, leftfield, field_level_eft, field_level_seljak, field_level_deep_learning, field_level_emulator}, since these approaches are shown to be even more sensitive to nonlinear and non-Gaussian signals in the data.
Therefore, it is crucial to explore approximation methods to produce mocks that accurately capture nonlinear effects to smaller scales (\(k > 0.2\,\hMpc\)) as this will ultimately determine how much information and physical insights we will be able to extract from upcoming surveys \citep{Nguyen_2021, field_level_eft}.
A vital challenge in generating fast and realistic mock catalogues is to model the link between the matter density field and the resulting galaxies or their host halos, denoted as galaxy or halo bias.

The scope of this paper focuses on constructing dark matter (DM) halos since around \(80 \%\) of the total matter is made out of dark matter and hence dominates the dynamics.
These halo catalogues are well-founded intermediate data products that can then be transformed into galaxy catalogues through methods such as the `Halo Occupation Distribution' (e.g. \citealt{hod1, hod2, hod3, hod4, hod5, hod6, hod7, hod8}, henceforth HOD), `Subhalo Abundance Matching' (e.g. \citealt{sham2, sham3, sham4, sham5, sham6, sham7, sham8}, hereinafter SHAM), \texttt{ADDGALS} \citep{addgals}, or semi-analytic models (e.g. \citealt{santa_cruz1, santa_cruz2}).

In this work, we propose a generative and physics-informed neural network we call \PT (\textbf{P}hysical and \textbf{in}terpretable \textbf{ne}twork for \textbf{tr}ac\textbf{e}r \textbf{e}mulation) that emulates the halo bias and can sample realistic dark matter halo catalogues in a computationally efficient way.
The subsequent section will briefly review existing mock generation approaches and set the context for our method.
We will then continue to describe the setup of our model, how it is trained, and used to generate mock catalogues in \cref{sec:pinetree,sec:data_and_conf}.
This naturally leads to \cref{sec:results} where we discuss our results before we conclude with a summary and outline for follow-up projects and applications.

\section{Overview of mock generation frameworks}
\label{sec:halo_bias}

The first analytical approaches to predict dark matter halo distributions, namely the halo mass function, from DM density fields were pioneered by \citet{og_kaiser, peak_background_split} and later refined (e.g. through works by \citealt{pbs2, pbs3, pbs4, pbs5, pbs6, pbs7}) to the formalism known now as `peak background split'.
In contrast, an alternative approach based on `excursion sets' \citep[e.g.][]{ps2, ps3, ps4, ps5, pbs5, ps7, ps8, ps9} that was initially formulated by \citet{press_schechter74} as the Press-Schechter formula was developed.
It was later extended by \citet{extended_PS} and further improved with the addition of ellipsoidal collapse \citep{extended_PS2, extended_PS3}.
Using these physically derived schemes, it is possible as a first approximation to predict the halo mass function and, at first order, the linear relation between the halo and matter density contrast 
\begin{equation}
    \delta_{\mathrm{h}} = b\,\delta_{\mathrm{m}},
\end{equation}
with \(b\) as a scale-independent bias parameter.
\(\delta_m\) and \(\delta_h\) denote the (dark) matter and halo relative density contrast respectively.

However, the linear bias is only a valid approximation for the largest cosmological scales as it fails to account for any nonlinear \citep{nonlinear_bias} and scale-dependent \citep{scale_dependent_bias} behaviour.
\citet{local_bias_expansion} extended the analytical framework to quasi-linear scales through the `local bias expansion'.
The perturbative bias expansion expresses the halo number density (or galaxy number density) as a function of the matter density field and its higher-order derivatives like the tidal field.
For an extensive review of the perturbative bias treatment, we refer the reader to \citet{bias_review}.
Notably, the concept of `local-in-matter-density bias' (\citealt{bias_review}, hereinafter LIMD bias) has been extended to phenomenological bias models beyond the linear bias such as power law parametrisations \citep{powerlaw_bias, powerlaw_bias2} or other nonlinear bias functions \citep{truncatedpowerlawneyrinck2014, ising_bias}.
The main idea of using LIMD bias models for mock generation is to produce a mean halo density field from a smoothed matter density field and then sample the halo field (e.g. through a Poisson distribution) to get a discrete count field.

\begin{figure}[h!]
	\centering
	\resizebox{\hsize}{!}{\includegraphics{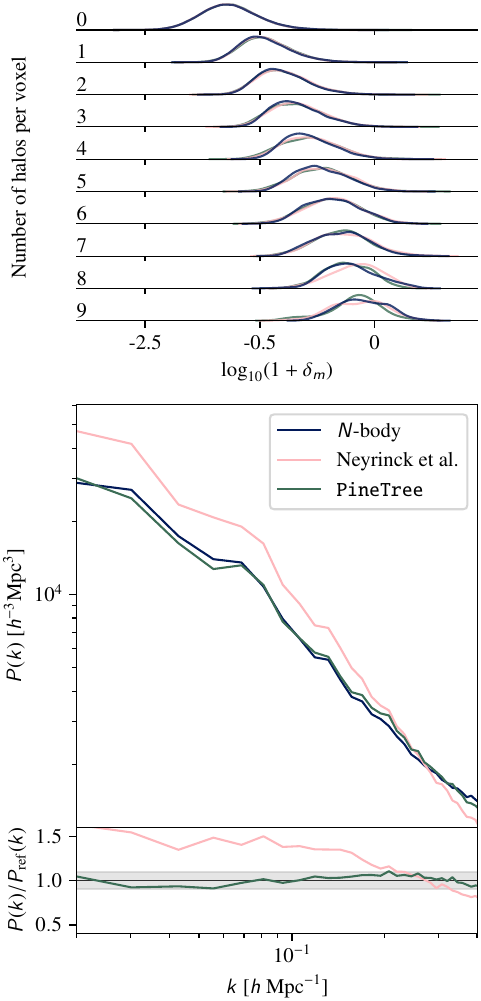}}
	\caption{Comparison between using one-point statistics (upper panel) and two-point statistics (lower panel) as validation metrics. The first panel on top shows a kernel density estimate plot of the number of voxels with a given halo count as a function of the underlying dark matter overdensity. Each line corresponds to the number of halos per voxel as indicated by the annotated text on the left. The distributions were produced using a  Gaussian kernel with bandwidth set according to Scott's rule \citep{scotts_rule}. The dark blue, pink, and green curves show the number distribution obtained from $N$-body simulation, truncated power-law by \citet{truncatedpowerlawneyrinck2014}, and \PT respectively. In the bottom panel, we show the mean power spectrum from the generated fields and ratios with respect to the reference \(N\)-body simulation. The grey band indicates the $10\%$ deviation threshold.}
	\label{fig:tpl_npe_comparison}
\end{figure}

\citet{non_local_bias, non_local_bias2}, and highlighted again by \citet{non_local_bias_needed} show that the local bias is insufficient to predict the halo distribution correctly.
Here, it is also important to emphasise that phenomenological bias methods need to be validated thoroughly since there is no theoretical guarantee of their accuracy, unlike the perturbative approach.
\Cref{fig:tpl_npe_comparison} depicts the breakdown of a local biasing scheme as well as the discrepancies in different validation metrics for which we fit the truncated power law (TPL) bias to a gravity-only \(N\)-body simulation following \citep{truncatedpowerlawneyrinck2014}.
We show that the calibrated TPL can reproduce the overall halo count distribution of the simulation, effectively validating the one-point statistics.
But when comparing the power spectra (two-point correlation) we can see a clear difference between the sampled mock fields and the reference \(N\)-body simulation.
\citet{non_local_bias_needed} show that this trend is true for any LIMD bias scheme.
Therefore, it is imperative to validate the mocks across different statistical orders especially if the science target is to generate realistic halos on the field level.

So far we have only discussed methods utilising fast analytic halo bias schemes to generate mocks.
An alternative to reduce (considerable) computational cost is to accelerate the generation of the underlying dark matter density by approximate gravity solvers.
This type of approach also known as `Lagrangian methods' involves the computation of an approximate matter distribution coupled with a deterministic halo formation prescription (see \citealt{review_mock_generation} for a more comprehensive review).
The first methods of this class such as \texttt{PINOCCHIO} \citet{pinocchio, pinocchio_2, pinocchio_3}, \texttt{PEAK-PATCH} \citep{peak_patch}, and \texttt{PTHalos} \citep{pt_halos, pt_halos2, pt_halos3} adopted Lagrangian perturbation theory for the density field computation.
Later works propose to utilise more accurate gravity solvers based on the particle-mesh (PM) scheme such as \texttt{PMFAST} \citep{pmfast}, \texttt{FastPM} \citep{fastpm}, \texttt{COLA} \citep{cola, cola2, cola3, cola4}, and its variants \texttt{L-PICOLA} \citep{lpicola} and \texttt{sCOLA} \citep{scola, scola2}.
One downside of these algorithms is that they rely on Friends-of-friends or other particle-based halo finders which can result in a computationally intense process when high accuracy is required (see for instance \citealt{computational_comparison_blot, review_mock_generation}).
Still, these methods have been readily adopted in combination with HOD or SHAM to generate mock surveys as demonstrated by \citet{sdss_mocks, des_mocks, des_mocks2, des_y3_mocks}.

The natural extension of these approaches is to combine approximate gravity solvers to produce smoothed density fields with simplified bias prescriptions as done by \texttt{PATCHY} \citep{Patchy}, \texttt{QPM} \citep{qpm}, \texttt{EZmocks} \citep{ez_mocks}, and \texttt{Halogen} \citep{halogen}.
Notably, these methods include a stochastic component to address the stochasticity of halo bias \citep[as discussed by e.g.][]{halo_stochasticity1, halo_stochasticity2, halo_stochasticity3, halo_stochasticity4, halo_stochasticity5, halo_stochasticity6, halo_stochasticity7, halo_stochasticity8, halo_stochasticity9, halo_stochasticity10}.
Despite utilising (local) bias schemes that can be difficult to calibrate, this type of mock generation approach has been successfully applied by \citet{patchy_mock, patchy_mock2, mocks_with_stochastic_bias} to produce large suites of accurate mock catalogues (see also \citealt{nifty_compare_mocks, review_mock_generation} for a comparison of some of the methods).

In recent years new techniques for mock generation have been explored to achieve even greater accuracy and computational efficiency.
For instance \texttt{BAM} \citep{bam, bam2, bam2.5, bam3, bam4} proposes a non-parametric bias scheme that samples halos based on 
the underlying matter density field and its associated cosmic web-type information motivated by \citet{cosmic_web_dependence, cosmic_web_dependence2}.
Meanwhile, \citet{modi_fcn_mocks, dm2haloCNN, ml_mocks, pandey2023charm, hybrid_bias} have investigated the use of machine learning as bias emulators since deep neural networks as universal function approximators are powerful tools to accurately model the nonlinear behaviour in the small-scale regime.
Yet, due to their large and degenerate parameter space, neural networks can be difficult and computationally expensive to set up and train, requiring large amounts of high-fidelity training samples.
Moreover, large networks are black-box models that can not be easily interpreted after training.
The following section will introduce our mock generation pipeline \PT that leverages machine learning techniques to incorporate a stochastic and non-local bias scheme while maintaining an interpretable and reduced parameter space through a physics-motivated model architecture.

\section{\texttt{PineTree}}
\label{sec:pinetree}

\begin{figure*}[ht!]
	\centering
	\includegraphics[trim={0 3cm 0 6cm},clip,width=18cm]{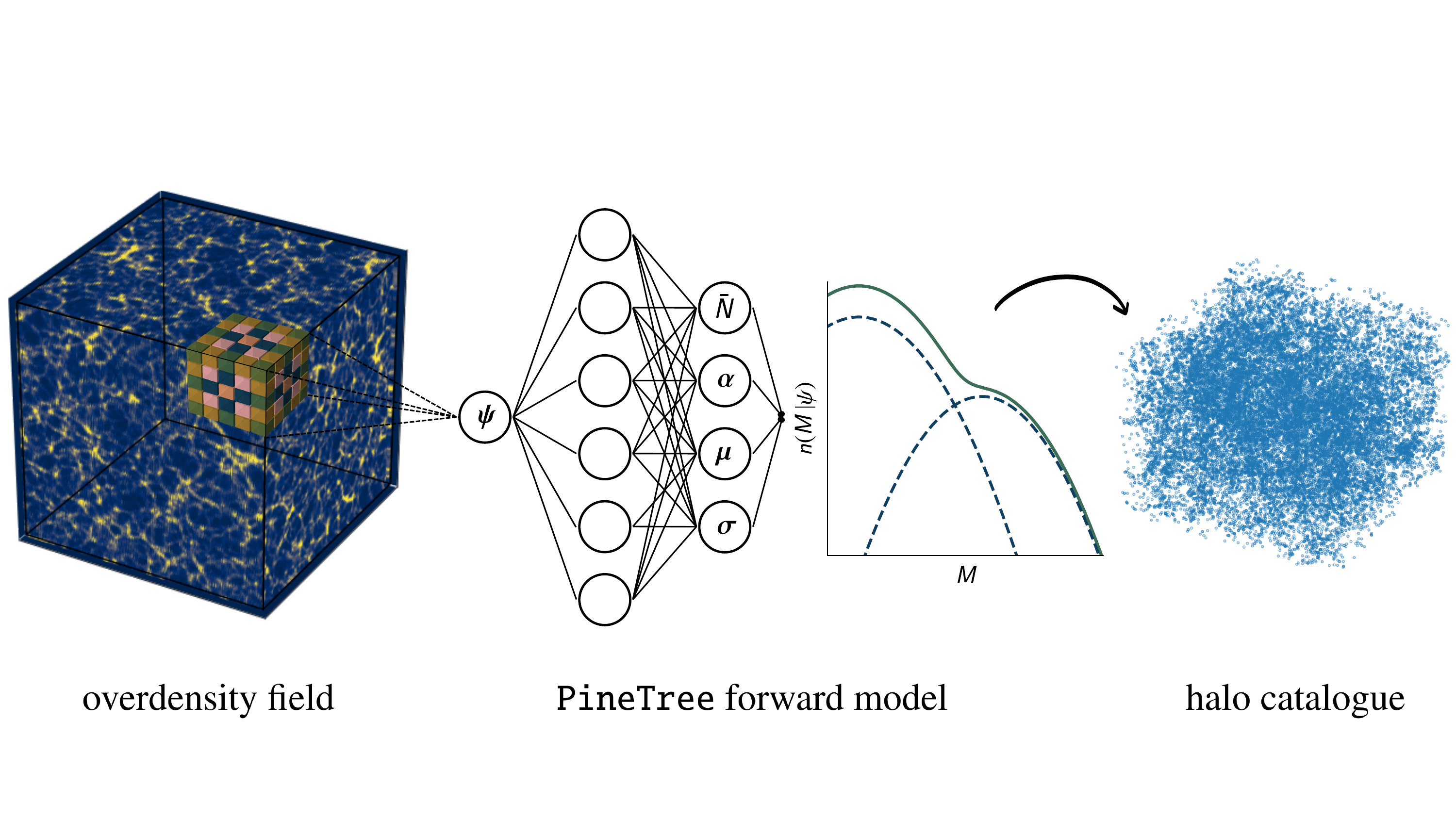}
	\caption{A schematic view of the \texttt{PineTree} pipeline.
	The following steps are depicted from left to right: The input dark matter overdensity field is convolved with a rotation-symmetric kernel; A mixture density network computes the distribution parameters of a log-normal mixture distribution with the feature map \(\psi\); Halo masses are drawn from the emulated distributions to construct the final halo catalogue.}
	\label{fig:pinetree_pipeline}
\end{figure*}

\texttt{PineTree} is a computationally fast and generative model with a minimal set of trainable parameters that takes as an input a gridded dark matter overdensity field $\delta_{\text{m}}(x)$ and outputs a halo count overdensity field $\delta_h(x)$ or halo catalogue \(\left\{ M_{h}(x) \right\} \) at the same gridding resolution.
The mock generation pipeline builds upon the Neural Physical Engine (NPE) first described by \citet{npe1} and is divided into three parts.
The first subsection presents the model architecture and how it differs from conventional machine learning approaches.
We then derive the objective function to optimise the model's parameters and conclude the section by describing how the trained model generates the final halo catalogues.
For an overview of the setup see the schematic outline in \cref{fig:pinetree_pipeline}.
The code is written in \texttt{Python} and is implemented in the \texttt{Jax}\footnote{\url{https://github.com/google/jax}} framework to enable \texttt{GPU}-acceleration.
A public version of the source code is hosted on Bitbucket\footnote{\url{https://bitbucket.org/dingzimeng/npe}}.
\subsection{Network architecture}
\label{subsec:network_architecture}

The network architecture consists of two functional parts.
A module to encode environmental information of the input and a neural network that approximates the halo mass function.
As mentioned in \cref{sec:halo_bias}, we want the network to consider each cell and its surrounding environment.
Thus, a network architecture building on a convolutional neural network (CNN) is the natural choice.
Furthermore, we want to encode our prior knowledge into the design of the network, namely that the principal physical forces of structure formation such as gravity are isotropic with regard to the underlying particle distribution.
This means that our bias model should a priorily neither be sensitive to the specific orientation nor the location of the dark matter overdensity field when predicting the resulting halo distribution.
Hence, the network operation performed on the input field needs to be rotationally and translationally invariant.

\subsubsection{Building environment summaries with a CNN}
\label{subsubsec:cnn}

A standard convolution layer \citep{cnn} performs a discrete convolution on an input field \(I\) with a set of kernel weights \(\omega^{\rm{Kernel}}\) and is therefore, by definition, already translation invariant.
In the three-dimensional case, the layer operation with a kernel of size \(M \times N \times L\) can be formally expressed as
\begin{equation}
    \psi_{ijk} = \left(I * \omega^{\rm{Kernel}}\right)_{ijk} = \sum_m\sum_n\sum_l I_{i+m,\,j+n,\,k+l}\,\omega^{\rm{Kernel}}_{mnl} + b,
    \label{eq:discrete_convolution}
\end{equation}
where the set of indices \(\{ijk\}\) relate to a given voxel in the discretised three-dimensional field.
In the above equation the index \(m\) runs from \(- (M-1)  / 2\) to \((M-1)  / 2\) and in the same way also for the indices \(n\) and \(l\) to center the convolutional kernel on the input voxel \(I_{ijk} \).
In the machine learning language, the output \(\psi_{ijk}\) is called a feature map and \(b\) is denoted as the neural bias parameter that needs to be learned alongside the kernel weights.
In the context of this work, however, the term bias will henceforth refer to the dark matter halo connection, unless stated otherwise.
The feature map \(\psi_{ijk}\) can be understood as the compressed information of the overdensity field patch around the central overdensity voxel \(\delta_{ijk}\).

\begin{figure}[h]
	\centering
	\resizebox{\hsize}{!}{\includegraphics{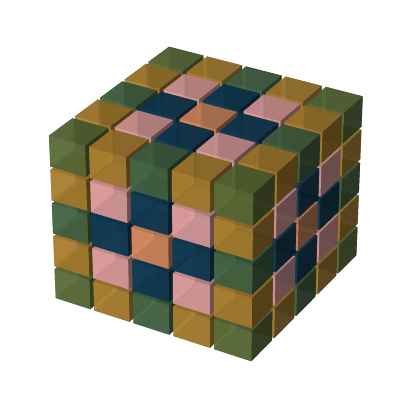}}
	\caption{Visualization of a $5 \times 5 \times 5$ monopole kernel. Identical colours indicate shared weights resulting in a rotationally symmetric pattern.}
	\label{fig:monopole_kernel}
\end{figure}

To enforce rotational invariance, we apply a multipole expansion to the convolutional kernel.
Inspired by a conventional multipole expansion where a generic function is decomposed as
\begin{equation}
	f(\theta, \varphi) = \sum_{l=0}^{\infty} \sum_{m=-l}^{l} C^m_l Y^m_l(\theta, \varphi),
\end{equation}
with $C^m_l$ denoting the constant coefficients and $Y^m_l$ the spherical harmonics, we construct a kernel that has shared weights for kernel positions with
\begin{equation}
	C^m_l(r,\theta, \varphi) = r \, Y^m_l(\theta, \varphi) \label{eq:shared_weight_criteria},
\end{equation}
where \(r\) is the Euclidean distance from the centre of the kernel to the position of the weight.
Effectively, we decompose a convolutional kernel with independent weights into a set of kernels with shared weights, for given \(l\) and \(m\), which is determined by the spherical harmonics, i.e., kernel positions with the same \(C^m_l(r, \theta, \varphi)\) according to \cref{eq:shared_weight_criteria}.
For a rotationally invariant kernel, we only consider the monopole contributions with $l=m=0$.
The output of the convolution operation then only encodes rotational and translational invariant features.
In addition, using this multipole expansion approach, the number of independent weights is greatly reduced. 
For instance, a $5 \times 5 \times 5$ kernel will have only 10 independent weights instead of $125$.
\Cref{fig:monopole_kernel} visualizes the shared weights of a monopole kernel.

To reduce the parameter space and maximise the computational speed of the model, we adopted a single filter, i.e., the monopole kernel, for the convolutional network.
Therefore the feature map \(\psi_i\) for a patch around a central overdensity voxel \(\delta_i\) is presented by a scalar.
We note that our public implementation of the method is already designed so that it is possible to include higher-order multipoles to handle direction-dependent effects (e.g. tidal forces) and potential improvements of utilising a multi-filter convolutional network approach will be explored in future works.
Each multipole contribution will result in a separate kernel, so in general, the output of the convolutional layer can also become a vector \(\boldsymbol{\psi}\) with each entry corresponding to a distinct feature map associated with a different kernel.
The relation to the multipole expansion gives us a direct way to interpret the kernel weights of our model, indicating how important specific spatial locations are for the model prediction.
In \cref{subsec:model_interpretation}, we will discuss the interpretation in detail for a trained monopole kernel.

\subsubsection{The mixture distribution}

The second part of the \PT model is designed to digest the non-local information encoded in $\psi$ and predict a probability function that can then be used to generate halo samples.
In this case, we want to model the conditional halo mass function \(n(M|\left\{\delta\right\}_{i})\), i.e. the number of expected halos for mass \(M\) conditioned on an overdensity patch \(\left\{\delta\right\}_i\) for the $i$-th voxel.
To ensure that the model can produce flexible and well-defined distributions, we chose a mixture of log-normal distributions, which are also frequently utilised in the machine learning community as density estimation networks.
The formal expression of the conditional halo number density then reads:
\begin{equation}
	n(M|\left\{\delta\right\}_{i}) = \frac{\bar{N}_i}{V}\sum_{n=1}^{\Nmix} \frac{\alpha_{ni}}{M \sqrt{2\pi\sigma_{ni}^{2}}} \exp\left[ - \frac{\left( \ln M - \mu_{ni} \right) ^{2}}{2\sigma_{ni}^{2}} \right],
	\label{eq:mdn}
\end{equation}
with \(\Nmix\) as the number of mixture distributions, $\bar{N}_i$ the predicted mean halo number for voxel $i$, and \(V\) denoting the voxel volume.
The conditioning on the overdensity patch enters through the previously computed feature map \(\psi_i\) via the mixture network parameters by:
\begin{align}
	\bar{N}_i &= \exp\left[\omega^{\bar{N}}\boldsymbol{\psi_i} + b^{\bar{N}} \right], \label{eq:n_bar} \\
	\alpha_{ni} &= \text{softmax}(\omega_{n}^\alpha\boldsymbol{\psi}_{i} + b_{n}^\alpha), \label{eq:amplitude_normalisation} \\
	\sigma_{ni} &= \exp\left[\omega_{n}^\sigma\boldsymbol{\psi_i} + b_{n}^\sigma\right], \label{eq:sigma}\\
	\mu_{ni} &= \begin{cases}
          \omega_{n}^\mu\boldsymbol{\psi_i} + b_{n}^\mu, &\text{if $n = 0$}, \\
          \text{max}\big[0,\ \omega_{n}^\mu\boldsymbol{\psi_i} + b_{n}^\mu \big] + \mu_{n-1}, &\text{if $ n > 0.$} \label{eq:mu}
          \end{cases}
\end{align}
Note that the parameters \(\left\{ \omega_{n}^{\bar{N}}, b_{n}^{\bar{N}}, \omega_{n}^{\alpha}, b_{n}^{\alpha}, \omega_{n}^{\mu}, b_{n}^{\mu}, \omega_{n}^{\sigma}, b_{n}^{\sigma} \right\} \) are defined globally for all voxels and that the density dependency only enters through \(\psi_{i}\).
Furthermore, the activation function in \cref{eq:amplitude_normalisation} ensures the mixture distribution is properly normalised so that \(\sum_n \alpha_{ni} = 1\) holds.

\subsubsection{Final transformations}

The final model is put together by first transforming the input overdensity \(\delta_m(x)\) into logarithmic space \(\log[1+\delta_m(x)]\) and then compressing it with the monopole kernel by a convolutional layer. 
The extracted feature map is subsequently transformed by the `softplus' activation function before finally being chained with the mixture density network.
Thus, the full network describes a non-local and nonlinear transformation from an overdensity environment patch to a halo number distribution at the environment centre.

It should be highlighted again that, due to the physically motivated design choices, the model only contains a very small parameter set compared to contemporary neural networks which typically have hundreds of thousands to millions of free parameters.
This is realised by utilising the symmetric single-kernel convolutional layer, and the shallow network that translates the feature map \(\psi\) to the mixture distribution parameters as described in \crefrange{eq:n_bar}{eq:mu}.
For a kernel with size \(3\times 3\times 3\) and a mixture density network of two components, the \PT model has a total of 18 parameters, making it suitable to directly infer the model parameters from a given dataset in the context of Bayesian inference or using it as a Bayesian neural network.
\Cref{subsec:model_interpretation} discusses the effect of utilising different kernel sizes for which the number of model parameters can increase.

\subsection{Loss function}
\label{subsec:loss_func}

The optimisation problem at hand is to find the best model parameters so that the sampled halo catalogues from our model match the halo catalogues obtained from a \(N\)-body halo finder.
The evaluation function quantifying the distance of the prediction to the ground truth is called the objective or loss function.
In line with the purpose of having an interpretable model, we use a physical likelihood function as the objective function following the maximum likelihood principle to optimise the model parameters.
This approach also enables the easy incorporation of \texttt{PineTree} into existing Bayesian forward modelling frameworks. 

The probability distribution of the observed data \(d\) given the underlying forward model with parameters \(s\) is referred to as the likelihood $\mathcal{P}(d\,|s)$.
The explicit form of the likelihood depends on our assumption of the stochasticity for the observable.
The likelihood, in the case of additive noise, can be formally written down by defining the measurement equation
\begin{equation}
	d = R(s) + n,
	\label{eq:measurement}
\end{equation}
where the function $R(s)$ denotes the forward model that maps the signal to the expected observation and $n$ denotes an additive noise.
Using \cref{eq:measurement} and assuming a reasonable distribution for the measurement noise $\calP_{\mathcal{N}}(n)$ (often Gaussian or Poissonian), the likelihood can then be written down as
\begin{align}
	\calP(d\,|s) &= \int \calP(d, n\,| s) \mathop{}\!\mathrm{d}n = \int \calP(n\,|s) \calP(d\,| n, s) \: \mathop{}\!\mathrm{d}n \nonumber \\
	       &= \int \calP(n\,|s) \delta_{D}(d - R(s)) \: \mathop{}\!\mathrm{d}n \nonumber \\
	       &= \calP_{\mathcal{N}}\left(n=d-R(s)\right), \label{eq:likelihood} 
\end{align}
with \(\delta_{D}\) denoting the Dirac delta function.

In this work, we assume that the number of halos per mass bin and per voxel is Poisson distributed
\begin{equation}
	N^{\rm halo}_{i, M} \sim \calP_{\rm Poisson}(\lambda_{i, M}),
	\label{eq:poisson_distribution_assumption}
\end{equation}
where \(\lambda_{i, M}\) denotes the mean (i.e., the Poisson intensity), and the indices $i$ and $M$ refer to the \(i\)-th voxel from the grid and the mass bin $\left[ M, M+\Delta M \right]$, respectively.
Therefore the mean halo number per voxel for each halo mass bin can be related to the conditional halo number density by
\begin{equation}
	 \lambda_{i,M} \equiv \left\langle N_{i,M} \right\rangle = V \int_{{M}}^{{M+\Delta M}} {n(M|\delta_i)} \mathop{}\!\mathrm{d} {M},
	 \label{eq:poisson_assumption}
\end{equation}
with \(V\) denoting the voxel volume.
The Poisson log-likelihood for observing $N^{\rm{obs}}_{i,M}$ is then given by
\begin{equation}
	\ln\calP_{i,M}(d=N^{\rm{obs}}_{i,M}\,| \lambda_{i,M}) = N^{\rm{obs}}_{i,M} \ln \lambda_{i,M} - \ln N^{\rm{obs}}_{i,M}! - \lambda_{i,M}.
	\label{eq:log_poisson_lh}
\end{equation}

For the optimisation scheme, only the gradient with respect to $\lambda_{i, M}$ is relevant.
Therefore, terms depending only on $N^{\rm{obs}}_{i, M}$ can be discarded, and we will omit them from now on.
Furthermore, we assume that voxels as well as halo mass bins are conditionally independent, thus yielding the total log-likelihood:
\begin{equation}
	\ln\calP(N^{\rm{obs}}\,| \lambda) = \sum_{i \in \text{voxels}} \sum_{M \in \substack{\rm{mass} \\ \rm{bins}}}N^{\rm{obs}}_{i,M} \ln \lambda_{i,M} - \lambda_{i,M}.
	\label{eq:npe_likelihood}
\end{equation}
If the halo mass bin width is decreased to the limit \(\Delta M \to 0\), then the observed number of halos \(N^{\rm{obs}}_{i,M}\)  can either be one or zero, and the above sums can be rewritten by
\begin{equation}
	\ln\calP(N^{\rm{obs}}\,| \lambda) = \sum_{j \in \rm{halos}} \ln \lambda_{i(j), M(j)} - \sum_{i \in \rm{voxels}} \int_{M_{\rm{th}}}^{\infty}  \mathop{}\!\mathrm{d} {M} \lambda_{i,M},
	\label{eq:npe_modified_lh}
\end{equation}
where the indices \(i(j)\) and \(M(j)\) correspond to the voxel and mass bin of halo \(j\) respectively.
Moreover, \(M_{\rm{th}}\) denotes the lower mass threshold of the halos in the simulation.

In the optimisation context, following maximum likelihood, the loss function \(\mathcal{L}\) is given by the negative log-likelihood.
So by inserting \cref{eq:mdn} and \cref{eq:poisson_assumption} into \cref{eq:npe_modified_lh}, the final expression that needs to be minimised reads:

\begin{align}
	\mathcal{L} &= -\sum_{j \in \rm{halos}}^{} \ln\left( \sum_{n=1}^{\Nmix} \frac{\tilde{\alpha}_{ni(j)}}{\sqrt{2\pi\sigma_{ni(j)}^{2}}} \exp\left[ - \frac{\left( \ln M_{j} - \mu_{ni(j)} \right) ^{2}}{2\sigma_{ni(j)}^{2}} \right] \right) \nonumber \\
		       &\phantom{=} + \sum_{i \in \rm{voxels}}^{} \int_{M_{th}}^{\infty} \sum_{n=1}^{\Nmix} \frac{\tilde{\alpha}_{ni(j)}}{M \sqrt{2\pi\sigma_{ni(j)}^{2}}} \exp\left[ - \frac{\left( \ln M - \mu_{ni(j)} \right) ^{2}}{2\sigma_{ni(j)}^{2}} \right] \mathop{}\!\mathrm{d} {M} \nonumber \\
		       &= -\sum_{j \in \rm{halos}}^{} \ln\left( \sum_{n=1}^{\Nmix} \frac{\tilde{\alpha}_{ni(j)}}{\sqrt{2\pi\sigma_{ni(j)}^{2}}} \exp\left[ - \frac{\left( \ln M_{j} - \mu_{ni(j)} \right) ^{2}}{2\sigma_{ni(j)}^{2}} \right] \right) \nonumber \\
		       &\phantom{=} + \sum_{i \in \rm{voxel}}^{} \sum_{n=1}^{\Nmix} \frac{\tilde{\alpha}_{ni(j)}}{2} \rm{erfc}\left( \frac{\ln M_{th} - \mu_{ni(j)}}{\sqrt{2\sigma_{ni(j)}^{2}}} \right). \label{eq:final_log_likelihood}
\end{align}

\subsection{Sampling of halos}

After obtaining the model parameters with the derived loss function from \cref{eq:final_log_likelihood} and a suitable optimisation scheme as described in \cref{subsec:model_conf}, the \PT model can then be used to generate mock halo catalogues for a given overdensity field.
The sampling procedure is divided into two parts: First sampling the total number of halos per voxel and thereafter the mass realisation of each halo.
For the number count generation, we need to use the initial assumption that the halos are distributed according to a Poisson distribution (see Eq. \ref{eq:poisson_distribution_assumption}).
Utilising the additive property of the Poisson distribution and the previous assumption of statistically independent mass bins, we can directly sample the halo count for all mass bins of one voxel rather than separately sampling each mass bin.
With regards to the \PT model, the predicted Poisson intensity per voxel is computed by
\begin{equation}
	\lambda_{i} = \sum_{M \in \substack{\rm{mass} \\ \rm{bins}}}\lambda_{i,M} = \sum_{n=1}^{\Nmix} \frac{\tilde{\alpha}_{ni}}{2} \rm{erfc}\left( \frac{\ln M_{th} - \mu_{ni}}{\sqrt{2\sigma_{ni}^{2}}} \right),
	\label{eq:predicted_poisson_intensity}
\end{equation}
which we subsequently use to generate a number realisation \(N^{\rm halo}_{i} \curvearrowleft \calP_{\rm Poisson}(\lambda_{i}) \).
Finally, according to the sampled halo count for each voxel, the individual halo masses are drawn in parallel from the predicted mixture density
\begin{equation}
	P(M) = \sum_{n=1}^{\Nmix} \frac{\alpha_{ni}}{M \sqrt{2\pi\sigma_{ni}^{2}}} \exp\left[ - \frac{\left( \ln M - \mu_{ni} \right) ^{2}}{2\sigma_{ni}^{2}} \right]
	\label{eq:halo_mass_distribution}.
\end{equation}
We note that the above expression only differs from the conditional halo number density from \cref{eq:mdn} by a factor of \(\bar{N}_i\,V^{-1}\).

\section{Training data and configuration}
\label{sec:data_and_conf}

Following the theoretical description of our model in \cref{sec:pinetree}, we detail the technical setup and parameter choices that led to the results shown in the subsequent section.
We first describe the \(N\)-body simulations that served as training and validation data before ending the section by reporting the specific model and training configurations that were tested.

\subsection{Description of the simulation dataset}
\label{subsec:data}

The \(N\)-body simulations were obtained by running a non-public variant of \texttt{Gadget-2} \citep{gadget2}, namely \texttt{L-Gadget} which is a more computationally efficient version that only simulates collisionless matter.
The initial conditions for the \(N\)-body simulations were computed by the initial condition generator \texttt{GenIC} \citep{nishimichi2009modeling, nishimichi2010scale, valageas2011combining}.
We produced 60 dark-matter-only simulations with different random seeds for the initial conditions using Planck 2018 cosmology \citep{planck2018}.
To capture large-scale dynamics with good particle resolution for luminous red galaxy halos \citep[e.g.][]{lrg_halo_mass}, we ran each simulation with box size of \(500\Mpch\) and that contained \(512^{3}\) particles, which results in a particle mass resolution of \(8.16 \times 10^{10}\,\hMsolar\).
We have set \texttt{GenIC} to use second-order Lagrangian perturbation theory (2LPT) to evolve the initial conditions until redshift \(z=24\) and the successive \texttt{L-Gadget} evolution to \(z=0\) and \(z=1\) was executed using a gravity softening of \(0.05 \ell\) where \(\ell \equiv (V_{\rm{box}} / N_{\rm{part}})^{1/3} \) is the mean inter-particle separation with \(V_{\rm{box}}\) as the volume of the simulation and \(N_{\rm{part}}\) denoting the total number of particles.
Using the same initial condition seeds as for the \texttt{L-Gadget} simulations, another set of 60 snapshots for each redshift \(z \in \{0, 1\}\) was produced through \texttt{GenIC} only.
Henceforth, we refer to this second set of simulations as 2LPT simulations.
We use the Cloud-in-Cell (CIC) interpolation described in \citet{cic} to obtain the final overdensity fields that are used as input for \texttt{PineTree}.

We validated the simulations by comparing the power spectra to the theoretical prediction from the Boltzmann solver \texttt{CLASS} \citep[e.g.][]{class, class2} which is shown in \cref{appendix:sim_validation}.
Finally, we used \texttt{ROCKSTAR} with linking length \(b=0.28\) and minimum halo particle set to \(10\) particles to produce the corresponding halo catalogues.
In this work, we adopted the virial mass definition according to \citet{virial_mass_bryan} leading to a lower halo mass threshold of \(3\times 10^{12}\,\hMsolar\) for each catalogue.
Again, we checked the halo catalogues for potential inconsistencies by comparing the simulated halo mass function to theoretical predictions from \citet{tinker2008} (See \cref{appendix:sim_validation}).

\subsection{Model configuration}
\label{subsec:model_conf}

In this work, we aim to study the scope of \texttt{PineTree}'s applicability.
Thus, we conducted an extensive parameter study and summarised the different configuration choices in \cref{table:model_conf}.
\begin{table}
\caption{Range of tested \PT configurations}
\label{table:model_conf}
\centering
\begin{tabular}{l c}
\hline\hline 
Setting & Used configurations \\
\hline\noalign{\vskip 2pt}
kernel size & $3^3,5^3,7^3 $ \\
voxel size $[\Mpch]$ & $50, 25, 16, 8, 4$  \\
redshift & $0, 1$ \\
gravity model & \texttt{L-Gadget}, \texttt{2LPT} \\
\hline 
\end{tabular}
\end{table}

For all results presented in the following section, the number of log-normal distributions was always set to two.
This was motivated by the intuition on halo mass functions provided by \citet{press_schechter74, tinker2008} and insights from the set of N-body simulations (see \cref{subsec:data}).
In addition to the listed configurations in \cref{table:model_conf}, we tested further configurations, such as a different number of mixture components.
However, these runs did not significantly improve our results and a complete report on the specific settings with their outcome is given in \cref{app:mdn_test}.

We opted for the gradient-based optimiser \texttt{Adam} by \citet{adam_opt} to achieve the optimisation.
The learning rate was set to \(0.001\) for all results shown in this work.
Other values and various learning rate schedulers were also tested without notable improvements.
We considered that the network parameters converged when the loss computed on the set of validation simulations saturates.
Unless stated otherwise, the 60 simulations are split into a training set comprised of \(10\) simulations and the validation set was made up of \(30\) different simulations that the model did not see during training.
The remaining \(20\) simulations were used for testing and we will justify our choice of only utilising \(10\) simulations for the training procedure in \cref{subsec:computational_cost}.

\section{Results}
\label{sec:results}

To assess the quality of the \texttt{PineTree} predictions, we compare the sampled mock catalogues with the halo catalogues from the dedicated validation set generated by \texttt{L-Gadget} and \texttt{rockstar}.
As described in \cref{sec:halo_bias} and highlighted in \cref{fig:tpl_npe_comparison}, it is important to validate by comparing the results with different levels of \(N\)-point correlation functions.
The main metrics used in this work comprise the halo mass function \(n(M)\), power spectrum \(P(k)\), and reduced bispectrum \(Q_{k_1, k_2}(\theta)\) that probe the 1-point, 2-point, and 3-point correlation respectively.
The power spectrum and reduced bispectrum were computed using the package \texttt{Pylians3} \citep{pylians}.
We begin this section by reporting the model efficiency with different runs following \cref{table:model_conf}.
We then discuss the insights that can be gained by examining the trained model weights and conclude with a benchmark on computational time.

\subsection{Best-case reconstruction}
\label{subsec:best_fit}

The runs with \texttt{PineTree} in this subsection aim to determine the best possible configuration for emulation and set up a reference to compare with subsequent experiments.
Therefore, we start with the dark matter overdensities from \texttt{L-Gadget} at redshift \(z=0\) and perform a grid search over the different field resolutions according to \cref{table:model_conf}, but with a fixed kernel size of \(3\times3\times3\).

\begin{figure}
    \centering
	\resizebox{\hsize}{!}{\includegraphics{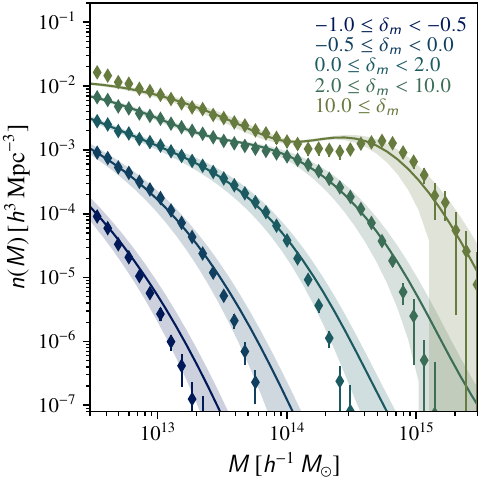}}
	\caption{The mean conditional halo mass functions as predicted by the \texttt{PineTree} model in solid lines and the halo mass function estimated from \(N\)-body simulations depicted in diamonds. The different colours correspond to different underlying dark matter overdensity bins that are specified by the legend in the figure. Error bars correspond to the \(1\,\sigma\) deviation and are presented as solid vertical lines or as shaded regions for the \(N\)-body estimates and \PT predictions respectively.}
	\label{fig:npe_hmfs}
\end{figure}

We find that the best-performing setting is for mock samples generated at the \(7.81 \Mpch\) voxel resolution and for halos with virial mass \(M_{\rm vir}\) between \(3 \times 10^{12} \hMsolar\) and \(1 \times 10^{13} \hMsolar\).
\begin{figure*}
	\centering
	\includegraphics[width=18cm]{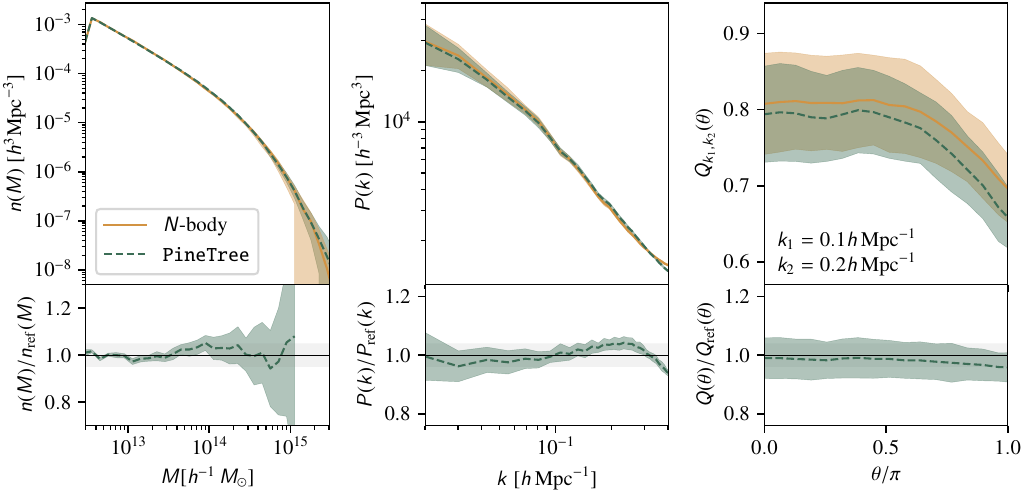}
	\caption{The marginal halo mass function, power spectrum, and reduced bispectrum from the left to right panel for halo density fields with virial mass between \(3\times 10^{12} \hMsolar\) and \(1\times 10^{13} \hMsolar\) at \(7.81 \Mpch\) voxel resolution.
     All curves are obtained from 30 catalogues with the dashed green and the solid orange line representing the ensemble mean of \texttt{PineTree} mocks and validation data respectively.
     The \(1\,\sigma\) line is outlined by the coloured shaded regions.
     The bottom subplot in each panel shows the ratio of the respective summary statistic between the mock and validation set where the shaded grey region is the \(5\%\) deviation threshold.
	} 
	\label{fig:best_fit}
\end{figure*}
\Cref{fig:best_fit} shows the three main metrics on the aforementioned best-configuration case.
For the marginal halo mass function shown in the first panel, we find good agreement between both catalogues to percent-level accuracy from the mass threshold up to halos of \(1 \times 10^{15} \hMsolar\), where the variance of the high mass halos starts to dominate.
Similar performance, with the two ensembles not deviating more than one sigma from each other, can be seen in the next two panels that display the comparison using the power spectrum and the reduced bispectrum. 
In addition, we report that the conditional halo mass function for a set of overdensity bins agrees between the predicted mock and validation catalogues as depicted in \cref{fig:npe_hmfs}.
We note that the shape of the predicted mass function by \texttt{PineTree} can change significantly over the range of underlying overdensities and we can see that the second log-normal distribution becomes apparent for high-overdensity regions.

\begin{figure}
	\centering
	\resizebox{\hsize}{!}{\includegraphics{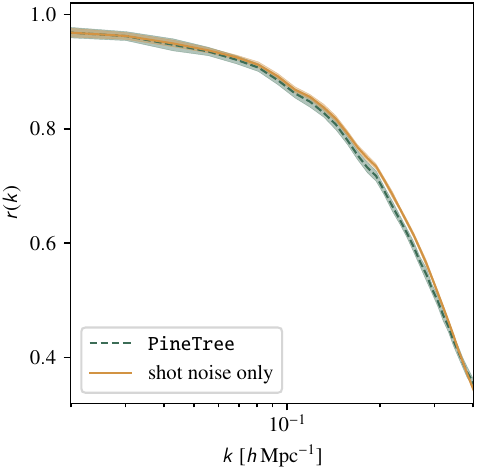}}
	\caption{Cross-correlation coefficient comparison for halos with masses between \(3\times 10^{12} \hMsolar\) and \(1\times 10^{13} \hMsolar\) at \(7.81 \Mpch\) voxel resolution from 30 validation samples. The green dashed line shows the mean for the cross-correlation coefficient from the mock fields while the mean of idealised fields with only Poisson noise contributions is depicted in solid orange. The shaded coloured area shows the \(1\,\sigma\) region respectively.} 
	\label{fig:cross_correlation}
\end{figure}
\begin{figure*}
\centering
   \includegraphics[width=17.8cm]{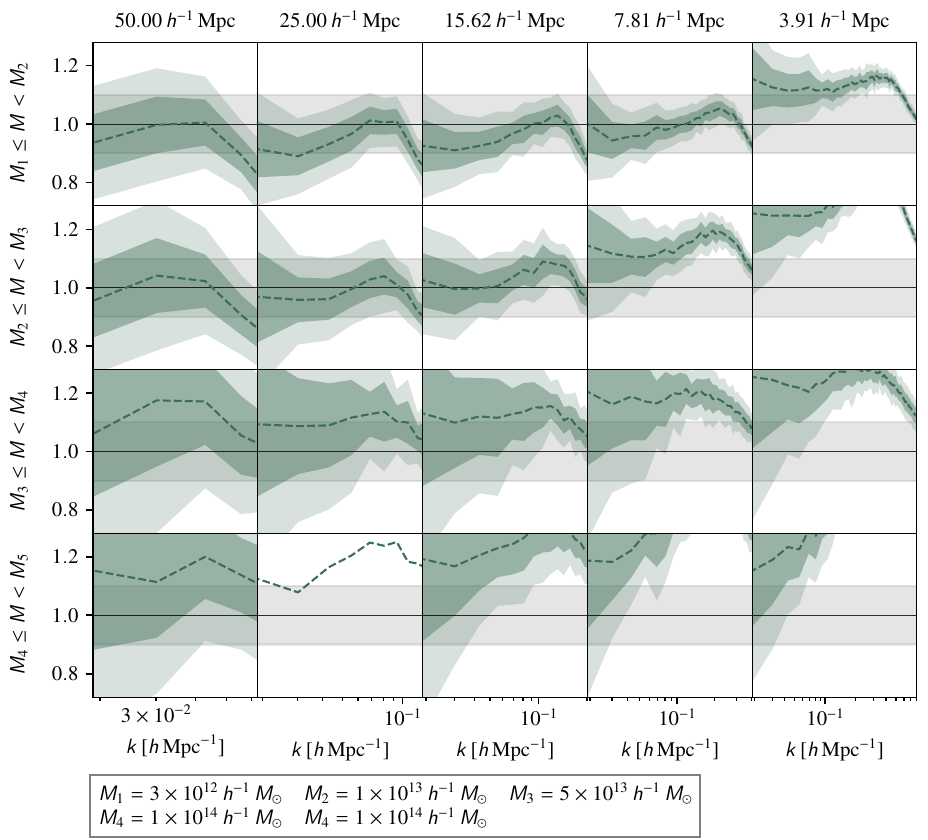}
   \caption{Ratios between the power spectrum of the \texttt{PineTree} mocks and the validation catalogues for different mass bins and at voxel resolutions. For the subplots shown from left to right, the voxel size is decreasing according to the \cref{table:model_conf} and also indicated by the annotations in the top row. From top to bottom, each row shows the ratio for increasing mass bins. As with all previous figures the quantities shown here are all computed using an ensemble of 30 fields.}
    \label{fig:overview_power_spec}
\end{figure*}
Computing the power spectrum for a given halo count field averages over the phases and hence, the comparison between the mock and validation catalogues yields only the difference in the auto-correlation strength per \(k\)-bin.
We want to further assess if the \texttt{PineTree} halos are realistic on the field level.
Therefore, to complement the validation at the two-point correlation level we compute the cross-correlation coefficient between each mock and validation halo count field via
\begin{equation}
	r_{\rm mr}(k) = \frac{P_{\rm mr}(k)}{\sqrt{P_{\rm mm}(k)P_{\rm rr}(k)}},
	\label{eq:cross_correlation_coefficient}
\end{equation}
with the indices `m' and `r' as labels for the mock and reference fields.
In \cref{fig:cross_correlation}, we show the cross-correlation coefficient computed for the 30 mocks and validation samples and we compare it to the best-case scenario where we assume that the decorrelation is only due to Poisson noise (also called shot noise). 
The cross-correlation coefficient with only shot noise contributions was obtained by assuming a given reference halo count field as ground truth and adding an uncorrelated Poisson noise \(\epsilon\) as a scale-independent term given by
\begin{equation}
	P_\epsilon(k) = \frac{V}{\bar{N}},
	\label{eq:poisson_noise}
\end{equation}
where \(V\) is the volume of the voxel and \(\bar{N}\) is the mean number of halos from the reference halo catalogue.
We can see that the actual cross-correlation coefficient traces the optimal case very closely concluding that the positions of the predicted halos correspond very well with the assumed ground truth from the full simulation.

\Cref{fig:overview_power_spec} shows the validation of \texttt{PineTree} mocks at power spectrum level for the different resolutions from \cref{table:model_conf} and across 4 mass bins.
We can see that the agreement with respect to the \(N\)-body ground truth deteriorates for higher mass bins as well as higher resolutions.
There are likely two main reasons for this, namely initial Poisson and conditionally independent voxel assumption.
Using a Poisson likelihood has the advantage that it can be written down in an analytic fashion [see \cref{eq:final_log_likelihood}], but as a consequence, the variance of the predicted halo distribution is always equal to the mean.
However, studies by works \citet{dm_halo_distribution_casas} indicate that for high overdensity regions and therefore, especially for high masses, the Poisson assumption breaks down.
The exact extent and impact of the non-Poisson contributions are not in the scope of this article and will be investigated in future works.
The second assumption of voxel independence is most likely to break down for high-resolution setups.
Because \texttt{PineTree} samples halos independently for each voxel the clustering behaviour of halos at small scales is neglected.
In other words, if a very massive halo is sampled in one voxel the neighbouring voxels will not be aware of this and the sampling behaviour is not adjusted.
This effect is then especially apparent at finer gridding since nearby voxels will tend to have similar underlying dark matter overdensities.
A more quantitative study on the impact of conditional dependencies with respect to grid resolution is planned and possibilities to address this issue will be discussed later in the conclusion section.
We saw a similar deterioration in agreement for the cross-correlation coefficient and the reduced bispectrum.

\subsection{Approximate gravity model}

\begin{figure}
	\centering
	\resizebox{\hsize}{!}{\includegraphics{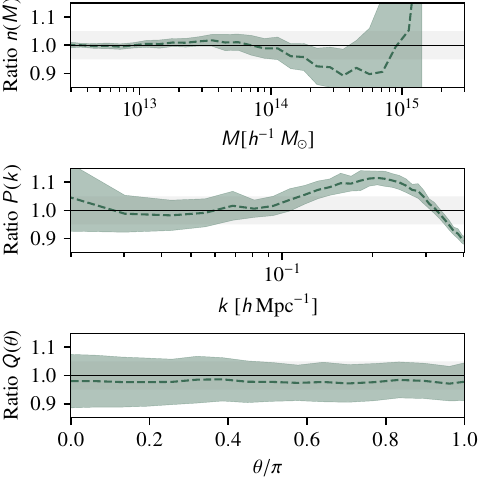}}
	\caption{Halo mass function, power spectrum, and reduced bispectrum ratios as shown in full by \cref{fig:best_fit} but for input overdensity fields from a \texttt{2LPT} gravity model. The respective quantities were computed for halos with a virial mass between \(3\times 10^{12} \hMsolar\) and \(1\times 10^{13} \hMsolar\) at \(7.81 \Mpch\) voxel resolution.}
	\label{fig:2lpt}
\end{figure}

The results in the previous subsection were predictions based on \texttt{L-Gadget} overdensity fields.
In practice, we would like to substitute the input fields with overdensity fields from approximate gravity solvers to save even more computational time.
\Cref{fig:2lpt} shows the results from \texttt{PineTree} retrained using the density fields produced by a \texttt{2LPT} model (as described in \cref{subsec:data}).
The figure is a simplified version of \cref{fig:best_fit} and shows only the ratios for the halo mass function, the power spectrum, and the reduced bispectrum with respect to the reference halo catalogue.
We can report that the precision of the prediction is almost identical to the setup with the \texttt{L-Gadget} overdensity fields showing that \texttt{PineTree} is able to interpolate correctly from the approximate dark matter fields to the high resolution \(N\)-body halos.

\subsection{Redshift dependency}

The variation discussed in this subsection is the application of \PT on \texttt{L-Gadget} dark matter overdensities and halo catalogues at redshift \(z = 1\) (as described in \cref{subsec:data}).
We find that the retrained model performs very similar to the setup at redshift \(z = 0\) with notably only a slightly worse agreement for the reduced bispectrum.
Therefore, we derive that \texttt{PineTree} can potentially be used to generate halo mocks at different redshifts.
Future studies of the model are planned to incorporate the redshift as a conditional parameter.
The results for this setup are shown in \cref{fig:z1_results}.

\subsection{Model interpretation}
\label{subsec:model_interpretation}

As mentioned in \cref{sec:pinetree}, the architecture of the network is interpretable and the trained weights can yield insights into the optimisation task itself.
In particular, we can draw intuition from the symmetric convolutional kernel on the effective environment size needed as an input for the network to predict the halo distribution for a given voxel.

\begin{figure*}
	\centering
	\includegraphics[width=18cm]{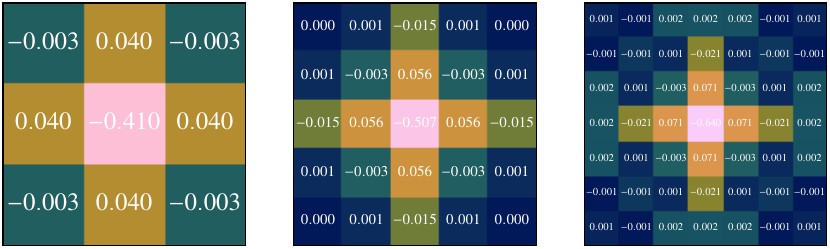}
	\caption{The trained kernel weights from the central slice through the monopole kernel of \texttt{PineTree}.
		 Each panel each from left to right shows a different kernel with size \(3^3\), \(5^3\), and \(7^3\) respectively.
		 The weights from the kernels are from optimisation runs at \(7.81 \Mpch\) voxel resolution and the underlying colour indicates the absolute value of each weight.
	 	}
	\label{fig:kernel_weights}
\end{figure*}

The kernel as described in \cref{subsubsec:cnn} is based on a multipole expansion and more specifically all setups use only the monopole.
\Cref{fig:kernel_weights} visualises a slice through each of the three kernels with varying sizes according to \cref{table:model_conf}.
All depicted kernels are from training runs where the fields are gridded with a voxel size of \(7.81 \Mpch\).
We see that the absolute values of the kernel weights strongly decrease the further it is located from the centre of the kernel.
Because the magnitude of each weight in a convolutional kernel indicates the relative importance of a given location of that patch, we can quantify the information content from an overdensity region needed to predict the subsequent halo properties.
We report that the central kernel itself accounts for over \(40 \%\) of the total kernel weight. 

\begin{figure}
	\centering
	\resizebox{\hsize}{!}{\includegraphics{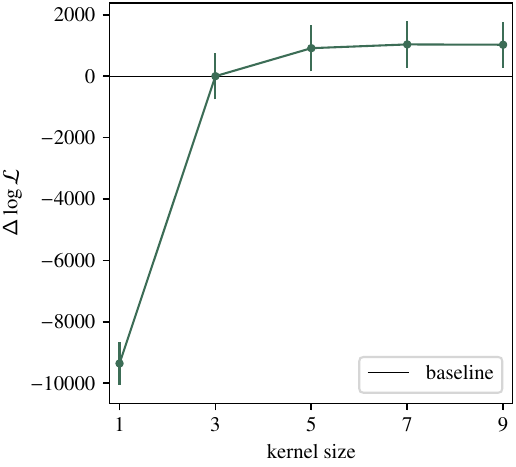}}
	\caption{The mean log-likelihood difference computed from the \(30\) validation simulations for \texttt{PineTree} with varying kernel sizes. The black horizontal line indicates the baseline model and represents the \(3\times3\times3\) kernel setup. Error bars for each point indicate the \(1\,\sigma\) deviation obtained from the validation set.}
	\label{fig:kernel_size_comparison}
\end{figure}

\begin{figure}[]
	\centering
	\resizebox{\hsize}{!}{\includegraphics{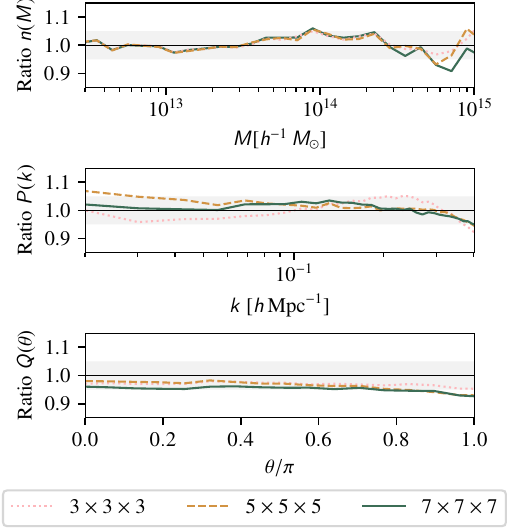}}
	\caption{Simplified version of \cref{fig:best_fit} showing the halo mass function, power spectrum, and reduced bispectrum ratios but for different kernel sizes as specified in \cref{table:model_conf}. The respective quantities were computed for halos with a virial mass between \(3\times 10^{12} \hMsolar\) and \(1\times 10^{13} \hMsolar\) at \(7.81 \Mpch\) voxel resolution.}
	\label{fig:kernel_comparison_val_metrics}
\end{figure}

Since the loss function for \PT is a likelihood, derived in \cref{eq:likelihood}, we can directly compare the setups with varying kernel sizes through the difference in log-likelihood computed on the validation set.
\Cref{fig:kernel_size_comparison} depicts the comparison of the difference in the log-likelihood \(\Delta\log\mathcal{L}\) for \(4\) different kernel sizes with the \(3 \times 3 \times 3\) kernel as baseline.
As expected, the local kernel performs the worst and there is a strong improvement when adopting the baseline model that takes neighbouring voxels into account.
Increasing the kernel size to \(5 \times 5 \times 5\) still yields a significant improvement in the log-likelihood while adopting even larger kernels does not improve the mean difference in log-likelihood beyond the standard deviation computed on the validation samples.
\Cref{fig:kernel_comparison_val_metrics} additionally visualises the performance across the \(N\)-point validation metrics.
For the \(9 \times 9 \times 9\) kernel, we observe over-fitting which would need to be mitigated via more training data or early stopping technique often utilised in the machine learning community, but this would digress from the goal of this work.
Therefore, we conclude that \texttt{PineTree} performs best in our resolution and parameterisation study for fields gridded at a voxel size of \(7.81 \Mpch\) and with a kernel size of \(7 \times 7 \times 7\).

\begin{figure}
	\centering
	\resizebox{\hsize}{!}{\includegraphics{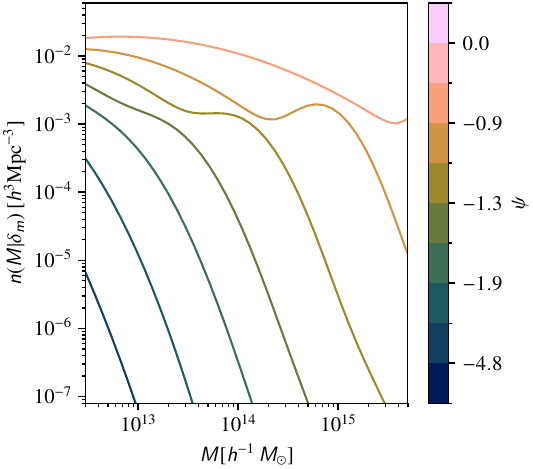}}
	\caption{Predicted conditional halo mass functions from \PT for different values (as indicated by the colour bar) of the feature map $\psi$. The scalar is as described in \cref{subsubsec:cnn} the output of the symmetric convolutional layer.} 
	\label{fig:mdn_vis}
\end{figure}

Another notable finding lies within the central \(3 \times\ 3\) patch of the kernel as shown in \cref{fig:kernel_weights}.
We can see that the signs between the central kernel weight and its adjacent weights are opposite to each other.
Consequently, a given matter density voxel and its surrounding environment patch adjacent positive overdensity regions will lower the contribution to the feature vector \(\psi\).
In contrast, voids with negative overdensity values will enhance the magnitude of the feature vector instead. 
\Cref{fig:mdn_vis} shows that with increasing \(\psi\) the resulting conditional halo mass function shifts right towards high mass halos and also increases in amplitude.
The result agrees with our physical intuition that extended overdensity regions yield less massive halos while regions where the overdensity is accumulated into a more condensed region will give rise to the most massive halos in our universe (or simulation).
Moreover, one can observe that the kernels in \cref{fig:kernel_weights} appear to be axis-aligned rather than purely isotropic which is due to the unphysical density estimation through the CIC kernel showing that \texttt{PineTree} is able to adjust for the alignment artefacts even with its constraint kernel setup. 
We show in \cref{app:sph_results} that the kernel weights are distributed more isotropically when using the computationally more expensive smoothed-particle hydrodynamics (SPH) interpolation \citep[first described by][]{sph, sph2}.

\subsection{Computational cost}
\label{subsec:computational_cost}

One of the key differences between our modelling approach and conventional machine learning methods is the significantly reduced number of parameters.
This leads to a model that is computationally efficient to train, both in terms of the required number of training samples as well as in computational time.
We report as an example the number of training simulations needed to saturate the validation loss of \texttt{PineTree} for the \(3^3\) kernel setup at \(7.81 \Mpch\) voxel size resolution.
The final mean validation log-likelihood computed for the \(30\) validation samples after training for a given training sample size are listed in \cref{table:training_sample_saturation}.
\begin{table}
\caption{Difference in log-likelihood depending on the number of training samples}
\label{table:training_sample_saturation}
\centering
\begin{tabular}{l c c}
\hline\hline 
Number of training samples & \(\mu(\Delta\log\mathcal{L})\) & \(\sigma(\Delta\log\mathcal{L})\) \\
\hline\noalign{\vskip 2pt}
1  & 0 & 753.02\\
5  & 9.05 & 753.58\\
10 & 10.48 & 753.08\\
20 & 10.84 & 753.09\\
30 & 10.75 & 753.18\\
\hline 
\end{tabular}
\end{table}
Again, we make use of the likelihood (see \cref{eq:likelihood}) to assess the training run via the difference in log-likelihood with the baseline model using only one simulation for training.
Already, the mean log-likelihood difference between one and five training simulations is significantly smaller than the standard deviation computed on the validation set and the difference becomes even more negligible when increasing the training set size.
This shows that \texttt{PineTree} can generalize well with a small number of training samples, and we were very conservative in our choice to train all our optimisation runs with the aforementioned \(10\) training simulations.

We conclude this section by reporting the computational time for training as well as running the model to produce halo catalogues to give an idea of the computational speed of our model.
The specific hardware used in this work was a 64-core AMD EPYC 7702 machine with 1 TB of RAM (hereafter referred to as the CPU machine) and a Nvidia V100 with 32 GB of RAM (denoted later on as the GPU machine).
The training was exclusively executed on the GPU machine, and training time mainly depended on the resolution on which the model was trained as reported in \cref{table:training_time} as it determined the total number of voxels for which the likelihood needs to be evaluated. 
For instance, at \(3.91 \Mpch\) voxel resolution, we need to compute over the likelihood for \(10 \times 128^3\) voxels as opposed to the \(50 \Mpch\) resolution run, where the training sample consisted only of \(10 \times 10^3\) voxels.
Due to the limited number of parameters, it would be possible to reduce the training time for high-resolution runs by only using sub-boxes of the training simulations, but we postpone a detailed study on optimising training time for \texttt{PineTree} for future work.
\begin{table}
\caption{Trainig time in GPU hours for different gridding resolutions}
\label{table:training_time}
\centering
\begin{tabular}{c c}
\hline\hline 
Number of GPU hours & voxel size \([\Mpch]\) \\
\hline\noalign{\vskip 2pt}
15  & 50\\
16  & 25\\
17  & 16\\
50  & 8 \\
100 & 4 \\
\hline 
\end{tabular}
\end{table}
Finally, we report the numbers for sampling halo catalogues using the trained model.
The sampling for a \(500\Mpch\) simulation gridded at \(128^3\) resolution took about $3.2$ seconds for the CPU and \(3.7\) seconds on the GPU.
This would result in an estimated cost of \(100\) GPU hours of training plus \(63\,000\) CPU hours to produce \(1\,000\) realisations of \(4 h^{-1}\,\mathrm{Gpc}\) sized mock halo simulations which is over an order of magnitude faster than \texttt{PINNOCHIO} (\(1\,024\,000\) CPU hours), \texttt{EZmocks} (\(822\,000\) CPU hours), or \texttt{PATCHY} (\(822\,000\) CPU hours) as reported by \citet{review_mock_generation}.

\section{Conclusions}
\label{sec:conclusion}

This paper has presented the computationally efficient neural network \texttt{PineTree} as a new approach to generating mock halo catalogues that are accurate on percent-level across various N-point correlation metrics.
We built upon the work of \citet{npe1} for which we further benchmarked and tested the limits of the model for a wide range of setups.

We performed an extensive resolution and parameter search to find that, in the current iteration, \PT performs best at a voxel resolution of \(7.81 \Mpch\) and for halo masses between \(3\times 10^{12} \hMsolar\) and \(1\times 10^{13} \hMsolar\).
Here we stress again that the metric utilised as the loss function during training is independent from the validation metrics, namely the halo mass function, power spectrum, cross-correlation, and bi-spectrum.
Therefore, this demonstrates the robustness of our model's predictions.
Similarly, the prediction accuracy translates well into setups with snapshots at redshift \(z=1\) and also for input density fields computed with a \texttt{2LPT} algorithm.
Notably, the use of \PT in conjunction with approximate density fields makes it attractive as an emulator for future works to generate large mock simulation suites that can then be utilized for instance to compute numerical covariances.

Subsequently, we discussed how the physics-informed model architecture facilitates interpretability.
The symmetric convolutional kernel and the simple network setup enable insights into the effective environment size needed to predict the halo masses optimally on a given gridding resolution.
In addition, the kernel reveals how the voxels in a density environment patch affect the resulting prediction and how much information each cell holds.
Furthermore, the physical Poisson likelihood employed as a loss function allows the seamless integration of \PT as the bias model into a galaxy clustering inference pipeline.
A natural follow-up study will leverage this alongside the model's reduced parameter space and differentiability within an inference with the algorithm \texttt{BORG} \citep{borg1, borg2, borg3, borg4, borg5}.

Finally, we showed in this work that \PT is not only computationally efficient for sampling halo catalogues but also for training.
Thus making the pipeline easily adaptable to different model setups without the need for large training data sets.

We see that the model's predictions start to break down for halo masses greater than \(1 \times 10^{14} \hMsolar\) and high gridding resolutions notably at \(3.91 \Mpch\).
The cause for this breakdown is related to the assumption of the underlying halo distribution.
First, dark matter halos follow a Poisson distribution and second, each voxel is independently distributed.
Future iterations of \PT can address these issues through the replacement of the loss and sampling procedure by a distribution-agnostic neural density estimator as done by \citet{pandey2023charm} or by incorporating a more flexible likelihood such as the generalised Poisson distribution \citep{consul1973generalization}.
To account for the halo conditional distribution across neighbouring voxels, a mass conservation scheme or multi-scale approach can be explored.

There are numerous potential applications of \PT to be tested.
Future extensions to this work can be the prediction of other halo properties like velocities or the direct generation of mock galaxy catalogues.
Another interesting aspect is the conditioning of \PT on cosmological parameters.

\begin{acknowledgements}

We thank Metin Ata, Deaglan Bartlett, Nai Boonkongkird, Ludvig Doeser, Matthew Ho, Axel Lapel, Lucas Makinen, Stuart McAlpine, and Stephen Stopyra for useful discussions related to this work.
We also thank Deaglan Bartlett, Ludvig Doeser, and Stuart McAlpine for their helpful feedback to improve the manuscript.
This work has made use of the Infinity cluster hosted by Institut d’Astrophysique de Paris. We thank the efforts of S. Rouberol for running the cluster smoothly. This work was granted
access to the HPC resources of Joliot Curie/TGCC (Très Grand Centre de Calcul) under the allocation AD010413589R1. 
This work was enabled by the research project grant ‘Understanding the Dynamic Universe’ funded by the Knut and Alice Wallenberg Foundation under Dnr KAW 2018.0067. JJ acknowledges support from the Swedish Research Council (VR) under the project 2020-05143 -- `Deciphering the Dynamics of Cosmic Structure'. GL and SD acknowledge the grant GCEuclid from `Centre National d’Etudes Spatiales' (CNES). This work was supported by the Simons Collaboration on `Learning the Universe'. This work is conducted within the Aquila Consortium (\url{https://aquila-consortium.org}). 

\end{acknowledgements}



\bibliographystyle{aa}
\bibliography{bib.bib} 



\begin{appendix}

\section{Simulation validation}
\label{appendix:sim_validation}

\begin{figure}
    \centering
    \resizebox{\hsize}{!}{\includegraphics{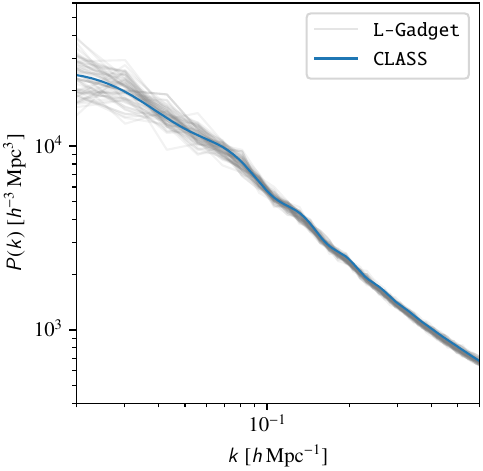}}
    \caption{The matter power spectrum computed from by the Boltzmann solver \texttt{CLASS} with nonlinear corrections at redshift \(z=0\) depicted by the solid blue line and the matter power spectra from the \(N\)-body simulations obtained from \texttt{L-Gadget} in light grey. Each grey line corresponds to one realisation from the simulation.}
    \label{fig:appendix_power_spec}
\end{figure}

\begin{figure}
    \centering
    \resizebox{\hsize}{!}{\includegraphics{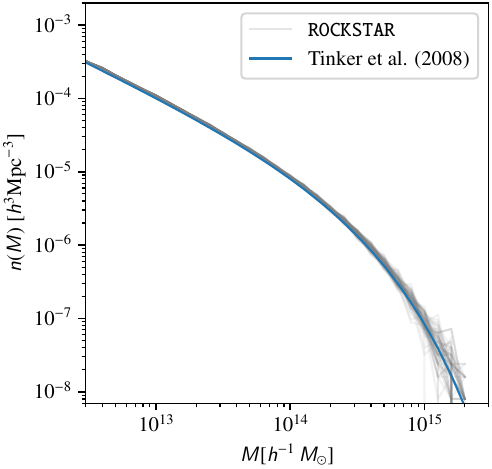}}
    \caption{Halo mass function comparison between \citet{tinker2008} (in solid blue) and the \(N\)-body simulations (in light grey) produced by \texttt{L-Gadget} and \texttt{ROCKSTAR} as halo finder. The halo mass function is for halos adopting the virial mass definition \citep{virial_mass_bryan} at redshift \(z=0\). Each grey line corresponds to one realisation from the simulation.}
    \label{fig:appendix_hmf}
\end{figure}

As described in \cref{subsec:data}, we compare the matter power spectrum and halo mass function with theoretical predictions to validate the simulations we generated with \texttt{L-Gadget} and \texttt{ROCKSTAR}.
\Cref{fig:appendix_power_spec} shows that the matter power spectra of the \(60\) simulations match the nonlinear theoretical prediction computed by \texttt{CLASS} \citep[as described in][]{class_nonlinear} very well for the Planck 2018 cosmological parameters \citep{planck2018}.
Similarly, the halo mass function reported by \citet{tinker2008} is in good agreement with the halo mass functions from the final \texttt{ROCKSTAR} halo catalogues as shown in \cref{fig:appendix_hmf}.
Therefore, we conclude that the simulation runs completed as expected and do not contain inherent errors that could bias our results in \cref{sec:results} obtained from the optimisations with \PT.

\section{Additional \PT optimisation runs}

\subsection{Mixture component testing}
\label{app:mdn_test}

\begin{figure}
    \centering
    \resizebox{\hsize}{!}{\includegraphics{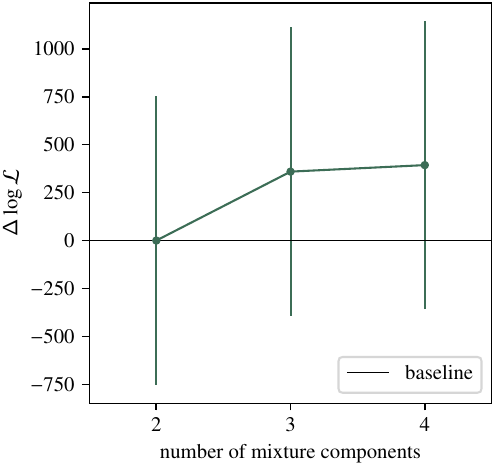}}
    \caption{The mean difference in log-likelihood computed from the \(30\) validation simulations for different numbers of mixture components. The black horizontal line indicates the baseline model and represents the \PT model with two mixture components. Error bars for each point indicate the \(1\,\sigma\) deviation obtained from the validation set.}
    \label{fig:appendix_mdn_val_loss}
\end{figure}

Following the description of \cref{subsec:model_conf}, we ran additional optimisations for different numbers of mixture components \(\Nmix\) in \PT namely two, three, and four.
Since the loss function is derived from a proper probability density (see \cref{subsec:loss_func}), we use the difference in the log-likelihood values computed on the validation simulations as the model comparison metric.
\Cref{fig:appendix_mdn_val_loss} depicts the log-likelihood difference \(\Delta\log\mathcal{L}\) for the three different runs with respect to the baseline model with two mixture components as used throughout \cref{sec:results}.
We optimised the different model setups on snapshots at redshift \(z=0\) and field discretisation at a voxel resolution of \(7.81 \Mpch\).
We report that even though the models with more mixture components slightly improve the log-likelihood compared to the baseline model, the improvements are not significant as the mean difference in log-likelihood is still within the \(1\,\sigma\) standard deviation of the baseline.
Hence, we limited our main resolution and parameter study to \PT configurations with two mixture components.

\subsection{Redshift \(z = 1\) results}
\label{app:redshift_z1}
\begin{figure*}
	\centering
	\includegraphics[width=17cm]{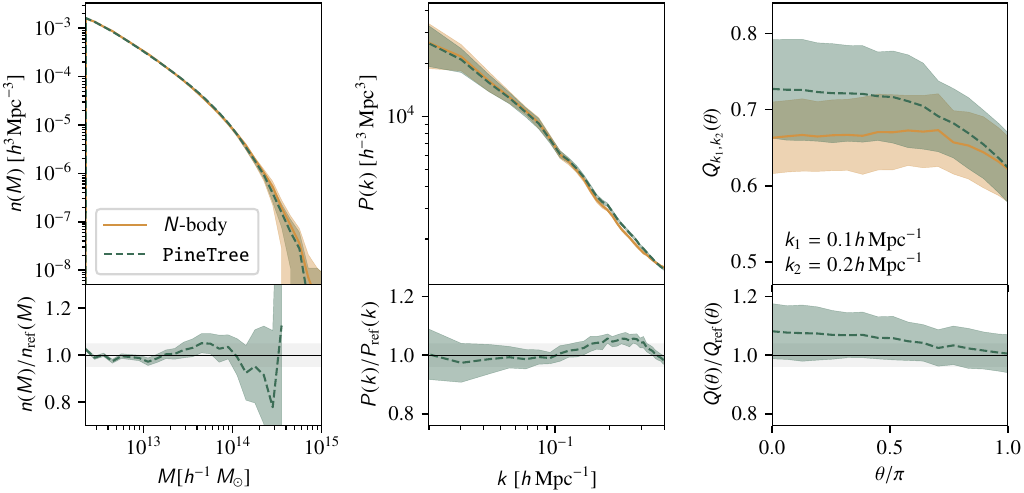}
	\caption{The same plots as shown in \cref{fig:best_fit}, but with halos at $z=1$. Again, the panels from left to right show the marginal halo mass distribution, the power spectrum and the reduced bispectrum with the network mocks plotted in green while the reference from the $N$-body simulations are shown in orange. The bottom subplot of each panel depicts the ratio of the respective summary statistics. The power spectrum and reduced bispectrum are computed using halos with a virial mass between \(3\times 10^{12} \hMsolar\) and \(1\times 10^{13} \hMsolar\).}
	\label{fig:z1_results}
\end{figure*}
Following the results of \cref{subsec:best_fit}, we optimise \PT with snapshots at redshift \(z=1\) at a voxel resolution of \(7.81 \Mpch\).
\Cref{fig:z1_results} shows the same three key validation metrics after training.
Comparing it to \cref{fig:best_fit} to the run at redshift \(z=0\), only the prediction for the reduced bispectrum is noticeably worse.

\subsection{Density fields from smoothed-particle hydrodynamics}
\label{app:sph_results}

As pointed out in \cref{subsec:model_interpretation} the optimisation with the overdensity fields computed via CIC leads to kernels that show axis-aligned effects (see \cref{fig:kernel_weights}).
To validate that this observed behaviour is not due to the model architecture itself, we retrain the same \PT model as described in \cref{subsec:best_fit} at \(7.81\Mpch\) voxel resolution, but with overdensity fields from \texttt{L-Gadget} estimated by the SPH algorithm.
We applied the SPH implementation developed and used by \citet{sph_implementation}.
Since the SPH interpolation scheme preserves the isotropic properties of the underlying particle distribution better than the CIC approach, we expect the change in the mass assignment scheme to be reflected by the resulting convolutional kernel.
\begin{figure}
	\centering
	\resizebox{\hsize}{!}{\includegraphics{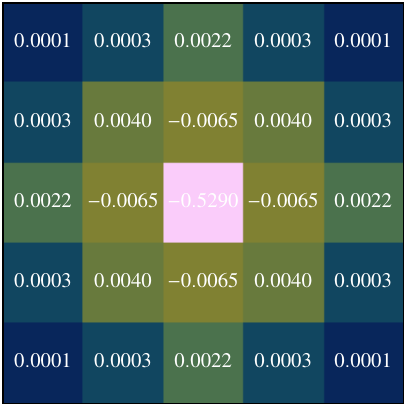}}
	\caption{The central slice through a  a \(5\times5\times5\) monopole kernel of \texttt{PineTree}, similar to \cref{fig:kernel_weights}. The weights are obtained through an optimisation run at \(7.81\Mpch\) voxel resolution and dark matter overdensity fields computed using an SPH algorithm. The underlying colour indicates the absolute value of each weight.}
	\label{fig:appendix_sph}
\end{figure}
\Cref{fig:appendix_sph} shows the trained kernel weights for the central slice. Comparing them to \cref{fig:kernel_weights}, we notice that the kernel weights appear more isotropic.
Hence, we conclude that the axis-aligned weights as shown in \cref{fig:kernel_weights} are caused by non-isotropic artefacts from the CIC interpolation.

\end{appendix}


\end{document}